\documentclass{aa}
\usepackage{graphics}
\usepackage{natbib}
\begin{document}
\title{Compact groups in the UZC  galaxy sample
\thanks{Tables 1 and 2 are only available in electronic at the CDS via
anonymous ftp to cdsarc.u-strasbg.fr (130.79.125.5) or via http://cdsweb.u-strasbg.fr/Abstract.html}}
\author{P. Focardi and B.Kelm} 
\offprints{P. Focardi, \email{Pfocardi@bo.astro.it}}
\institute{Dipartimento di Astronomia, Universit\`a di Bologna}
\date{Received 15 May 2001 / Accepted 1 March 2002}
\abstract{
Applying an automatic neighbour search algorithm to the 3D UZC galaxy 
catalogue \citep{Falco} we have identified 291 compact groups 
(CGs) with redshifts between 1000 and 10\,000 km\,s$^\mathrm{-1}$.  
The sample is analysed to investigate 
whether Triplets display kinematical and morphological characteristics similar to higher order CGs (Multiplets). 
It is found that Triplets constitute low velocity dispersion structures,
have a gas-rich galaxy population and are typically retrieved in sparse environments.  
Conversely Multiplets show higher velocity dispersion, include few gas-rich 
members and are generally embedded structures. 
Evidence hence emerges indicating that Triplets and Multiplets, though sharing a common scale, 
correspond to different galaxy systems. Triplets are typically field structures  
whilst Multiplets are mainly subclumps (either temporarily projected or collapsing) within 
larger structures. 
Simulations show that selection effects can only partially account 
for differences, but significant contamination of Triplets by
field galaxy interlopers could eventually induce the observed 
dependences on multiplicity.  
}
\maketitle
\section{Introduction}
Small galaxy systems such as pairs and Compact Groups (CGs) constitute 
the very lowest end  of the clustering hierarchical scale. 
Given their high galaxy density and small velocity dispersion most CGs 
are expected to separate from their underlying background, 
become bound systems and ultimately collapse within a few crossing 
times. Actually the high frequency (or extreme longevity) of CGs can match 
the rather short lifetimes predicted by merger simulations \citep{Barnes}  
simply by varying the fraction of dark matter distributed through the group 
\citep{Mamon87,Ath,Zabludoff}, assuming continuous accretion of infalling 
galaxies \citep{Governato}, or assuming that CGs are dense configurations 
that form temporarily within loose groups \citep{Diaferio}. 
An alternative scenario requires that 
merging CGs are continuously replaced by new forming ones \citep{Mamon00}. 

To date it is difficult to further constrain the relative importance of 
parameters and correlations entering the modelling of CGs, 
essentially because no definite conclusions concerning fundamental 
properties of CGs have been achieved. 
A large unbiased sample is needed to provide 
statistically reliable answers to questions such as: Do  
isolated CGs really exist? And how does the request for minimum 
multiplicity depend upon magnitude and morphological 
classification of member galaxies? Hence, questions 
related to a proper choice of CG selection parameters become fundamental, 
whilst actually, these parameters are generally 
chosen according to criteria aiming at reducing contamination by 
non-physical structures. 
  
Indeed, the bound status of CGs is difficult to establish. 
CGs, unlike galaxy clusters, though presenting adequate mass density 
profiles, are generally too close (z$<$0.1) to induce efficient gravitational 
lensing phenomena \citep{Mendes,Montoya},  
while concerning X-ray properties, diffuse emission 
tends to be associated only with embedded CGs in loose configurations that 
contain at least one early-type galaxy \citep{Ponman,Mulchaey,Heldson}. 
Other tracers of a common potential well, such as HI or CO, are at present 
available only for a limited number of 
CGs \citep{Williams,Oosterloo,Verdes-Montenegro}. 

Therefore proximity in projected and redshift space, although 
affected by small number statistics, peculiar motions and interlopers 
\citep{Moore,Diaferio}, 
still remains the main tracer of physical association between 
galaxies in CGs. Interaction patterns and kinematical peculiarities 
between member galaxies 
constitute an {\it a posteriori} probe of physical association.   

Because of the difficulty in identifying high redshift CGs, only low redshift 
CG samples are so far available. 
The best studied CG sample (HCGs, Hickson 1982,1997) contains 
92 CGs showing extremely heterogeneous characteristics. HCGs have been 
visually selected (according to multiplicity, isolation and luminosity 
concordance of member galaxies) and thus reflect some of the systematic 
biases intrinsic to identification of systems on the basis of 
their bidimensional distribution only. 
In order to overcome these biases automatic identification of CGs has been 
performed on a deep 2-D southern catalogue (SCGs,  
Prandoni et al. 1994; Iovino et al. 1999) and on 3-D catalogs 
(RSCGs, Barton et al. 1996, 1998).  
These studies intended to produce large CG samples by (partial) 
parametrical reproduction of Hickson's selection criteria. 
Hickson's isolation criterion  has been slightly relaxed by Iovino 
et al. (1999) and not included at all by Barton et al. (1996), 
who additionally, included triplets among CGs.  
Triplets are structures generally excluded in bidimensional selected CG  
samples because, apart from the expected high contamination 
by superposed fore/background galaxies, they might represent 
a collection of unrelated field galaxies, rather than a physical 
structure \citep{Diaferio}. 
The Catalogue of Triple Galaxies \citep{Karachentseva,Karachentseva2000} 
constitutes the exception, but because 
of poor number statistics affecting dynamical parameters, Triplets have 
so far been investigated mainly in relation to 
their high peculiar galaxy content. 
 
Recent availability of a 3-D large galaxy sample, including 
nearly 20\,000 redshifts for northern galaxies brighter than $m_B$=15.5 
(UZC, Falco et al. 1999), allowed us to construct a large CG sample 
selected on the basis of their compactness only.
In selecting the sample we did not try to reproduce any of Hickson's 
criteria except compactness, in order to check if 
and at which level the properties of CGs are linked to 
multiplicity, to the large scale 
environment and to the luminosity and spectral properties of member galaxies. 

The CG selection algorithm is described in Sect.\,2. In 
Sect.\,3 we describe the UZC catalogue and the 
prescriptions for the algorithm input parameters.   
The analysis of the characteristics of Triplets (Ts) and Multiplets  
(Ms) are presented in Sect.\,4. 
In Sect.\,5 statistical reliability of the CG sample 
is discussed.   
In Sect.\,6 spectral properties of CG galaxy members 
are presented. 
CGs large scale environment and surface density contrast 
are analysed in Sects.\,7 and 8 respectively. 
In Sect.\,9 the relation between adopted selection parameters and 
the properties of the resulting CG sample are discussed. 
Conclusions are drawn in Sect.\,10.   

A Hubble constant of 
$H_\mathrm{0}$=100\,{\it $h^\mathrm{-1}$}km\,s$^\mathrm{-1}$Mpc$^\mathrm{-1}$ 
is used throughout. 
\section{The CG identification algorithm }
In order to safely deal with CG multiplicity and properly compare T 
and M characteristics we have devised a CG identification algorithm 
imposing compactness as the only requirement.
The algorithm counts neighbours to each galaxy in 
3D space within a volume defined by projected distance $\Delta$r and 
velocity ''distance''
$\Delta$v$^\mathrm{I}$ ($\Delta$r and $\Delta$v$^\mathrm{I}$ 
are free input parameters). 
When a galaxy is found to have at least two neighbours the geometrical 
center of the system is identified. Additional members within $\Delta$r 
and $\Delta$v$^\mathrm{I}$ of the centroid are then searched for and a 
new center computed. 
This is an iterative process that goes on until convergence  
is reached, i.e. no further CG member is detected and no 
previously identified CG member gets excluded. 
Non-convergent systems are obviously rejected.  
CG centers are not weighted by magnitude of member galaxies on purpose, 
in order to enable non-biased investigation of possible relationships 
linking CG kinematics to luminosity.    

The searching method is asymmetric and may produce different grouping 
depending on which galaxy is selected first. 
In order to overcome this undesirable effect the algorithm retains in the 
main sample only CGs whose single galaxies all have no further neighbour 
(within  $\Delta$r and $\Delta$v$^\mathrm{I}$) 
except those already listed as members. 
Non-symmetric CGs are excluded, because without 
definition of further selection criteria, the algorithm is unable to 
define which galaxies are CG members and which are to be left outside.  
The symmetrization procedure also ensures that no overlapping 
CGs are retained.  
Finally, cross correlation with ACO clusters \citep{Struble} enables 
the algorithm to exclude from the sample CGs which are cluster substructures 
at distance less than 1\,R$_\mathrm{Abell}$ from the ACO centers. 

For each CG the local surrounding galaxy density is computed   
within the free input parameters $\Delta$R and $\Delta$v$^\mathrm{II}$. 
The algorithm also provides parameters indicative of 
average compactness and maximum physical 
extensions. These are the unbiased 
line of sight  velocity dispersion {\bf $\sigma$$_\mathrm{v}$}, 
the maximum difference in redshift 
space between a CG member and the center {\bf $\Delta$v$_\mathrm{max}$}, 
the radius {\bf r$_\mathrm{ave}$} measuring projected average galaxy distance 
from 
the center, and the radius {\bf r$_\mathrm{max}$} defined as the projected 
separation between the center and the most distant CG member galaxy. 
Average projected dimension of CGs (r$_\mathrm{ave}$) is preferred 
to the median value, because having imposed a maximum physical extension 
to CGs, each galaxy distance should be equally weighted.  

Our  algorithm displays some analogies and differences 
with the friends of friends (FoF) group searching algorithm 
by Huchra \& Geller (1982) and with the hierarchical procedure applied by 
Tully (1987). 
Like Tully (1987) our CGs are defined by internal conditions only and 
our procedure starts hierarchically by requiring 
a minimum galaxy density threshold to identify a CG. 
At variance with the FoF method, requiring a maximum galaxy-galaxy separation
as a function of redshift, 
we impose a maximum size for the CGs. Adopting a common scale for structures 
allows to safely deal with multiplicity but induces a redshift luminosity 
dependence. To correct for this bias the CG sample is divided in 4 
distance classes (see Sect. 3) and the comparison of CGs of different 
multiplicity is performed within each class. 
Moreover while the FoF procedure,
to discriminate between physical and non physical systems, 
 requires a minimum density contrast threshold 
(computed with respect to the average galaxy density of the sample), 
our CGs are identified without a constraint on density contrast.  
Instead, we do compute the surface density contrast locally  
(within $\Delta$R and $\Delta$v$^\mathrm{II}$) after CGs 
have been identified. 
The advantage of this approach is that we can perform non 
biased analysis of CG environments. 

\section{ The CG sample }
 
We have applied the CG searching algorithm to the UZC catalog 
\citep{Falco}, which is the widest angle   
redshift compilation available for nearby galaxies. 
The UZC catalog is a revised version of ZNCAT \citep{Tonry} which
was created for the first CFA redshift survey \citep{Davis}. ZNCAT is 
the union of CGCG galaxies in the Zwicky catalog \citep{Zwicky}  
and UGC galaxies in the Uppsala General Catalog \citep{Nilson}.
Inclusion within ZNCAT of UGC galaxies, which are selected applying also  
a diameter criterion,  
reduces the partial loss of low surface brightness extended galaxies in CGCG. 
UZC includes only ZNCAT galaxies with $m_B$$\leq$ 15.5, the limit at which
Zwicky estimated that his catalog was complete. 
The uncertainty on UZC galaxy magnitude is 0.3 mag \citep{Bothun,Huchra76}. 
UZC covers the entire northern sky down to declination $\approx$ 
-2.5$\degr$ and has no fixed limit on minimum galactic latitude. 
It is claimed \citep{Falco} to be 96\% complete for galaxies 
brighter than $m_\mathrm{B}$=15.5. 
The solid angle covered by UZC is $\approx$ 1.4$\pi$ sr. 
(for galaxies with $|$b$_\mathrm{II}$$|$ $\geq$ 20$\degr$). 

The CG sample we present here is specifically designed to allow 
comparison between compact Triplets and higher order CGs. 
Therefore, we have chosen to set {\bf $\Delta$r=200$h^\mathrm{-1}$kpc} and 
{\bf $\Delta$v$^\mathrm{I}$=$\pm$1000\,km\,s$^\mathrm{-1}$}. 
The prescription for $\Delta$r accounts for possible huge dark 
haloes tied to bright galaxies \citep{Zaritsky,Bahcall}. 
The value for $\Delta$v$^\mathrm{I}$ is large enough to allow a  
fair sampling of the CG velocity dispersion, which can be related to other   
observational parameters such as morphological content and 
surrounding galaxy density \citep{Somerville, Marzke}. 
Actually, more than 95\% of the CGs display $\sigma$$_\mathrm{v}$ values 
below 500 km\,s$^\mathrm{-1}$. 

Concerning the large scale, we have set 
{\bf $\Delta$R\,=\,1$h^\mathrm{-1}$Mpc} and 
{\bf $\Delta$v$^\mathrm{II}$=$\pm$1000\,km\,s$^\mathrm{-1}$} in order 
to map the environment on scales typical of loose groups/poor clusters. 
Moreover, adopting the same value for $\Delta$v$^\mathrm{I}$ and 
$\Delta$v$^\mathrm{II}$ ensures that each CG is sampled  
to the same depth of its large scale environment. 
Only CGs in a redshift range 1000\,km\,s$^\mathrm{-1}$ to 10\,000 
km\,s$^\mathrm{-1}$ enter the sample. 
The low redshift threshold allows us to reduce uncertainties due to 
peculiar motions, the upper one to reduce the incidence of CGs with 
only extremely bright galaxies. 

The search algorithm, applied to the UZC sample with the prescriptions just 
defined, yields a sample of 291 CGs: 
222 Triplets (Ts) and 69 Multiplets (Ms) with more than 3 member galaxies. 
The algorithm additionally detected (and rejected) 
56 ACO subclumps and 144 non-symmetric CGs, among which Ms are at 
least  50\%. 
The CG sample is shown in  Table\,1 which lists RA and Dec of the center 
(col.\,2 and 3), number of members n (multiplicity) 
(col.\,4), average projected dimension r$_\mathrm{ave}$(col.\,5), 
mean radial velocity cz (col.\,6),  
unbiased radial velocity dispersion $\sigma$$_\mathrm{v}$ (col.\,7) 
and, for CGs with cz $\geq$ 1500 km\,s$^\mathrm{-1}$ (see Sect.\,7), 
the number of large scale neighbours N$_\mathrm{env}$ 
within R=1$h^\mathrm{-1}$Mpc (col.\,8). 
Cross identification with HCGs and RSCGs is reported in col.9. 
 Table\,2 lists member galaxies for each CG, their position, magnitude, 
radial velocity and spectral classification as reported in UZC. 
The CG sample characteristics are shown in  Fig.\,1. 
\begin{figure}
\resizebox{\hsize}{!}{\includegraphics{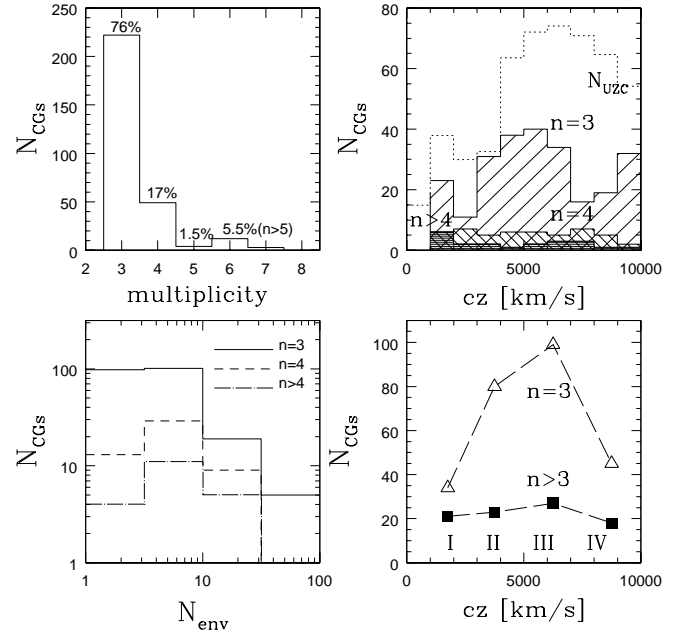}}
\hfill
\caption[]{In the upper left panel the CG distribution as a function of 
multiplicity is shown. The upper right panel shows redshift 
distribution for CGs with different multiplicity: Ts constitute 3/4   
of the CG sample. The dotted line represents, on an arbitrary scale, 
the distribution of UZC galaxies. The lower left panel shows the number of 
large scale neighbours (N$_\mathrm{env}$) for Ts, for CGs with 4 members, 
and for CGs with more than 4 members respectively. Ts constitute the 
majority of structures with few neighbours. 
In the lower right panel the number of Ts and Ms is plotted, 
within each of the 4 distance classes in which separate data comparison is 
performed. }
\end{figure} 

The CG distribution 
as a function of multiplicity (upper left panel) shows that Ts 
represent the majority of the sample. 
The upper right panel shows how the redshift 
distribution of CGs of different multiplicity compares to redshift 
distribution of UZC galaxies. 
The lower left panel shows the relation between CG multiplicity and 
the number of large scale neighbours N$_\mathrm{env}$. A correlation between 
multiplicity and large scale environment clearly emerges, with Ts representing 
the majority of the structures with few neighbours.
The KS test indicates that distributions between Ts and higher multiplicity 
CGs are different 
with significance level larger than 99.7\%.  

To extract physical information from the complete flux limited sample  
the role of the luminosity of member galaxies has to be properly 
disentangled, hence nearby CGs have to be separated from more distant ones. 
With this aim the sample was split into 4 distance classes whose radial velocities span over a 3000 km/s range each, with an overlap 
among adjacent samples of 500 km/s. 
The first subsample is actually slightly smaller because 
all CGs at redshift below 1000 km/s are excluded, and its 
overlap with the next subsample slightly larger. 
The 4 subsamples 
lie within 1000-3000 km\,s$^\mathrm{-1}$, 
2000-5000 km\,s$^\mathrm{-1}$, 4500-7500 km\,s$^\mathrm{-1}$, and 7000-10\,000 
km\,s$^\mathrm{-1}$ respectively (henceforth referred as 
subsamples I,\, II,\, III and IV). 
Subsamples mimic homogeneous samples, complete in magnitude and 
volume, and allow to correctly take into account multiplicity and neighbour density. 
The small overlap in redshift space 
does not bias the statistical analysis of the sample, as only 
Ts and Ms within the same subsample are compared, and no comparison 
between CGs in different subsamples is performed. 
 Table\,3 reports for each subsample the median value of the kinematical 
parameters provided by the algorithm, 
together with the median value of the large scale neighbours. 
The distribution of Ts and Ms, in the four defined distance classes, 
is shown in the lower right panel in Fig.\,1. 
The decline in both distributions in subsample IV reflects 
the sharply decreasing luminosity function of galaxies at the high 
luminosity end.
The fraction of  UZC galaxies in CGs within each of the 4 defined 
subsamples is 11\%, 10\%, 7\% and 4\% respectively.  
Actually, since the volumes covered are extremely different, our results on  
the 4 subsamples exhibit different levels of statistical significance. 
Subsample I should strongly reflect our position within the Local 
Supercluster. For example, several CGs in subsample I are Virgo cluster 
subclumps (see Mamon 1989).

The volume number density of all CGs  (computed for systems 
at cz$\ge$2500 km\,s$^\mathrm{-1}$ and $|$b$_\mathrm{II}$$|$$\geq$40$\degr$)  
turns out to be 1.6$\times$10$^\mathrm{-4}$$h^\mathrm{3}$Mpc$^\mathrm{-3}$, 
almost 4 times the density of Ms alone. CGs number density 
slightly exceeds values estimated in RSCGs \citep{Barton96}, which in turn, 
retrieve number densities much higher than in HCGs 
because of Hickson's bias against Ts. 

\setcounter{table}{2}
\begin{table*}
\caption[]{CGs kinematical and environmental parameters. The CG sample has been split  
according to redshift range. Figures for Ts and Ms are shown separately. 
Median values are tabulated.}
\begin{center}
\begin{tabular}{|l||rrrrrr||rrrrrr|}
\noalign{\smallskip}
\hline
\hline
\noalign{\smallskip}
subsample& {\bf Ts} &$\sigma$$_\mathrm{v}$& $\Delta$v$_\mathrm{max}$& r$_\mathrm{ave}$&r$_\mathrm{max}$&N$_\mathrm{env}$
&{\bf Ms} &$\sigma$$_\mathrm{v}$ & $\Delta$v$_\mathrm{max}$ & r$_\mathrm{ave}$&r$_\mathrm{max}$&N$_\mathrm{env}$ \cr 
   &  &[km/s] & [km/s] & [kpc] & [kpc] & - & & [km/s] & [km/s] & [kpc] & [kpc] & -\cr
\hline
\noalign{\smallskip}
I  & 34  &142 &162  &  77 &108  &8 & 21  &174 &216  & 76  &127 &8 \\
II & 80  &128 &141  &  74 &99   &3 & 23  &245 &279  & 82  &135 &5 \\
III& 99  &152 &175  &  65 &93   &3 & 27  &262 &380  & 82  &121 &4 \\
IV & 45  &174 &200  &  79 &109  &2 & 18  &314 &446  & 91  &122 &3 \\
\noalign{\smallskip}
\hline
\hline
\end{tabular}
\end{center}
\end{table*}

\section{The main properties of CGs} 
One essential question is whether Ts constitute a fair subsample 
of CGs, especially since Ts are much more numerous than Ms in all 
subsamples. 
With this aim, we compare here the magnitude of member galaxies and the CG 
kinematical-dynamical properties.   
The spectral properties and large scale environment are 
examined in section 6 and 7.    
As the KS test shows that, except for subsample I, Ms are more 
luminous than Ts (at 99.5\% c.l.) we first check whether 
a similar difference is also retrieved between the luminosity of 
Ts and Ms member galaxies. 
It is found that within each of the 4 distance classes 
Ms and Ts member galaxies display similar  
absolute magnitude distributions. 
Hence the larger luminosity associated with Ms simply arises 
from the fact that Ms include more members than Ts and does not indicate 
that higher multiplicity CGs are typically associated with brighter galaxies. 

\begin{figure}
\resizebox{\hsize}{!}{\includegraphics{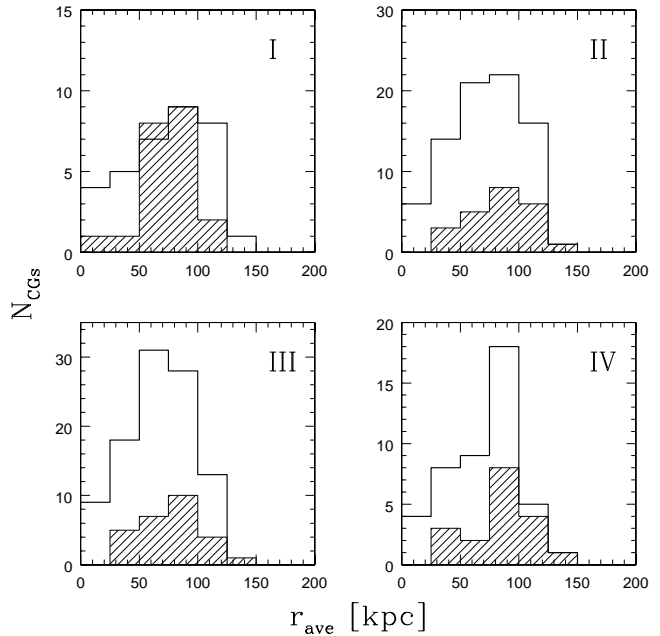}}
\hfill
\caption[]{Distribution of Ts and Ms 
(hatched) as a function of the parameter r$_\mathrm{ave}$ 
tracing the average projected dimension of the group.}
\end{figure} \begin{figure}
\resizebox{\hsize}{!}{\includegraphics{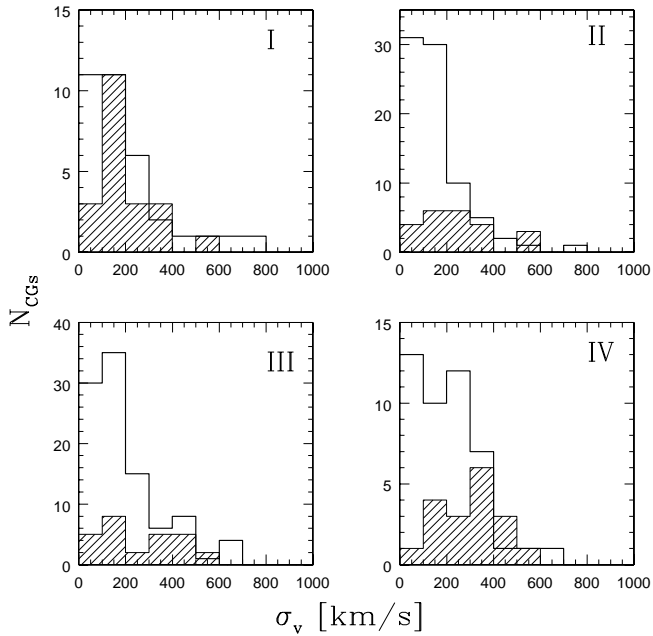}}
\hfill
\caption[]{Distribution of Ts and Ms (hatched) 
as a function of $\sigma$$_\mathrm{v}$, 
tracing the extension of the CG in redshift space.}
\end{figure} 

\begin{figure}
\resizebox{\hsize}{!}{\includegraphics{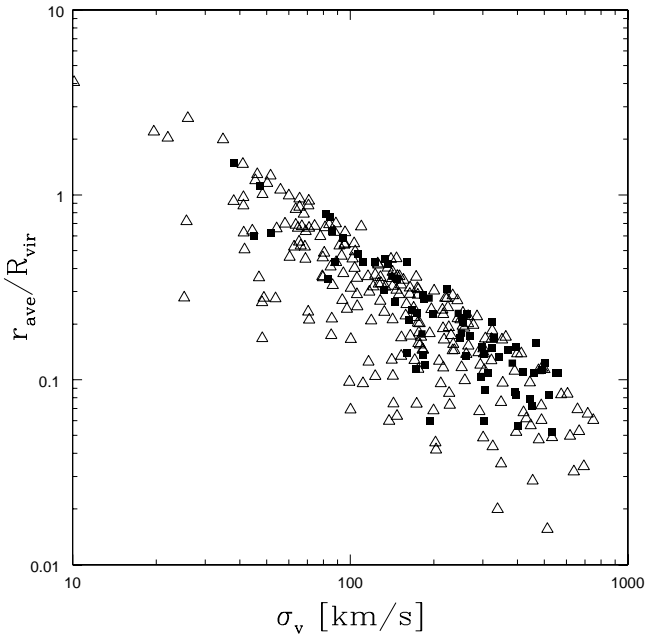}}
\hfill
\caption[]{Ratio of average dimension r$_\mathrm{ave}$ over 
virialization radius R$_\mathrm{vir}$, 
as a function of the parameter $\sigma$$_\mathrm{v}$ for 
Ts (empty triangles) and Ms (filled squares).} 
\end{figure} 

\begin{figure}
\resizebox{\hsize}{!}{\includegraphics{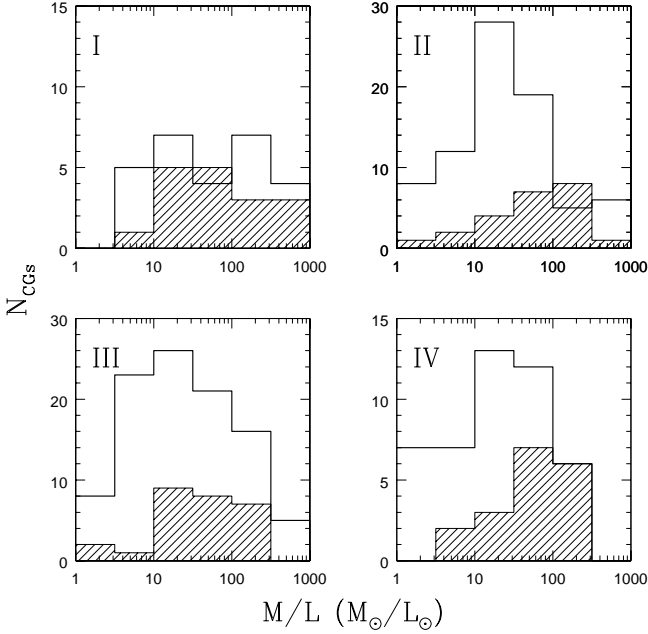}}
\hfill
\caption[]{$\mathrm{M/L}$ distribution of Ts and Ms 
(hatched). Ms display larger M/L than Ts at 46\%, 98\%, 87\% and
94\% c.l. in the four classes. }
\end{figure} 
\begin{figure}
\resizebox{\hsize}{!}{\includegraphics{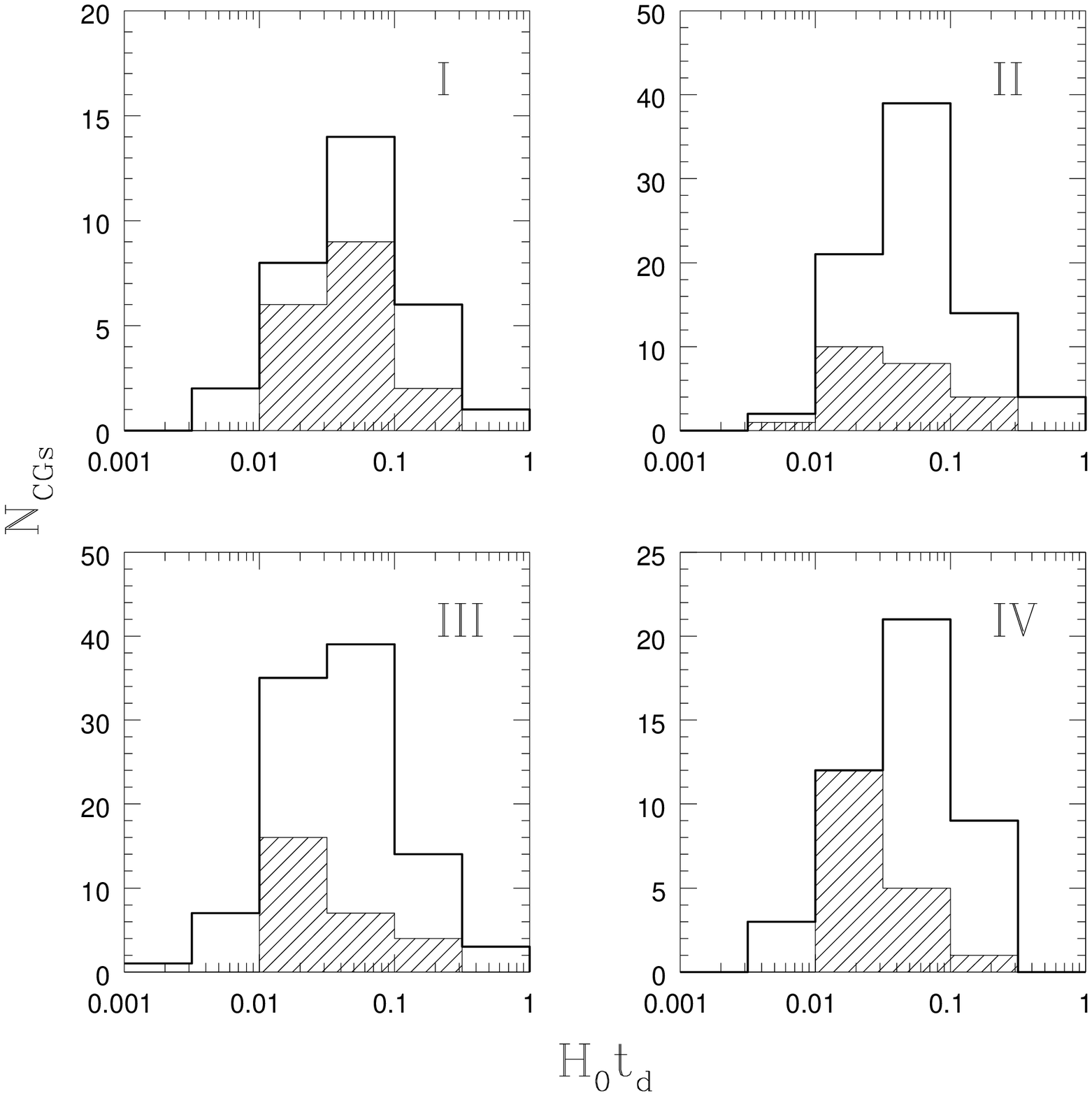}}
\hfill
\caption[]{Distribution of Ts and Ms (hatched) 
as a function of the dynamical time $H_\mathrm{0}$t$_\mathrm{d}$ in Hubble
time units. Ms display shorter $H_\mathrm{0}$t$_\mathrm{d}$ than Ts at 
42\%, 86\%, 90\% and 98\% c.l. in the four classes. }
\end{figure} 

As far as kinematical properties are concerned, Fig.\,2 shows the 
distribution of the average extension (r$_\mathrm{ave}$) for  
Ts (solid histogram) and Ms (hatched) in the four subsamples. 
Ts appear more compact than Ms in all but the first subsample. 
However, according to the KS test, differences in r$_\mathrm{ave}$ between 
Ts and Ms are not significant (59\%, 56\% and 77\% c.l. respectively). 
This is not unexpected, given our selection criteria, 
and actually confirms that we sample Ts and Ms on a common scale. 
When r$_\mathrm{max}$ rather than r$_\mathrm{ave}$ is examined 
differences get significant (above 90\% c.l.) in subsamples II and III.  
While $\approx$40\% of the Ms include a member which is 
at a distance larger than 150$h^\mathrm{-1}$kpc from the center,  
this is the case for less than 7\% of Ts. 
The excess of Ms with members close to the 
limiting distance, together with the high fraction of 
Ms among rejected non symmetric CGs, possibly indicates that we are 
sampling subclumps embedded 
in larger structures eventhough the external limit of 200$h^\mathrm{-1}$kpc 
imposed by the algorithm prevents from drawing definite 
conclusions concerning any typical dimensions for Ms. 
In the cz range between 2500 and 7500 km/s, including 
60\% of  Ms, 
the average dimension of CGs increases with multiplicity following 
the relation r$_\mathrm{ave}$ $\propto$ n$^\mathrm{0.6}$. 
This relation has been derived 
for the median number of galaxies in multiplets which is 4.5.   
   
The velocity dispersion of galaxies in a bound system provides 
an estimate of the potential well, although in CGs errors caused by random 
orientation of the system along the line of sight might dominate the result. 
In any case values obtained on a large sample of CGs are less affected by 
this bias, and thus yield more reliable results. 
In  Fig.\,3 the distributions of Ts and Ms relative to the parameter 
$\sigma$$_\mathrm{v}$ are shown. Distributions are different at  
61\%, 99.6\%, 97\% and 98\% c.l respectively. 
Comparison of $\sigma$$_\mathrm{max}$ yields obviously more significant 
differences (98\%, 99.99\%, 99.9\% and 99.8\% c.l.).   
Considering again CGs within the range 2500-7500 km/s, we find 
$\sigma$$_\mathrm{v}$ to increase with multiplicity as n$^\mathrm{1.4}$.  

Next, before estimating the mass associated with CGs, 
we check whether and how many CGs in the sample satisfy the 
necessary (but not sufficient) criterion 
for a galaxy system to be virialized.   
In  Fig.\,4 \,r$_\mathrm{ave}$/R$_\mathrm{vir}$ as a function of 
$\sigma$$_\mathrm{v}$ for Ts and Ms is plotted. 
R$_\mathrm{vir}$ is computed according to prescriptions in 
$\Lambda$CDM ($\Omega$$_\mathrm{M}$=0.3, $\Omega$$_\mathrm{\Lambda}$=0.7) 
cosmologies, requiring a virialized system to 
display an overdensity greater than 333 with respect to the mean density 
of the universe.  
Figure 4 shows that most CGs (95\%) in the 
sample satisfy the virialization condition and might therefore be 
physical bound systems. Had we compared R$_\mathrm{vir}$ 
with the harmonic radius, the fraction of virialized systems 
would be slightly lower (90\%).  

Concerning the real nature of CGs it must also 
be stressed that the median velocity 
dispersion associated with galaxies in Ts (Table 3) is comparable to the 
mean galaxy-galaxy velocity difference associated with field galaxies 
\citep{Somerville,Fisher}. Accordingly one could speculate that 
the Ts sample suffers from serious contamination by pseudo-structures
 of unrelated field galaxies (filaments viewed nearly edge on), 
in which redshift tracing the Hubble flow is used to compute 
a velocity dispersion. If this is the case 
the contamination by interlopers is expected to bias the velocity 
dispersion of Ts towards the low end. 
However the exclusion of suspiciously low-$\sigma$ systems 
would also cause any genuine bound CG  representing a system in its 
final state of coalescence to be excluded from the sample. 
In our sample the fraction of low $\sigma$$_\mathrm{v}$ CGs  
(i.e.  $\sigma$$_\mathrm{v}$$\leq$100 km\,s$^\mathrm{-1}$) turns out to be
 32\% and 16\% among Ts and Ms.  
The first value is slightly lower than the 
40\% unbound Triplets claimed by Diaferio (2000).  
Figures are roughly  
consistent given that Diaferio selects systems with a FoF algorithm, 
which, when applied to small systems, tends to return an excess of 
elongated structures displaying enhanced contamination by outliers. 
Concerning Ms, the bias induced by interlopers might well 
push the velocity dispersion higher so that it is not 
obvious how to separate structures contaminated by interlopers from 
bound structures. 

The substantial difference in the kinematical characteristics 
of Ts and Ms 
might affect also parameters directly derived from 
$\sigma$$_\mathrm{v}$ and r$_\mathrm{ave}$ such as estimated mass 
(M $\propto$ $\sigma$$^2_\mathrm{v}$$\times$r$_\mathrm{ave}$) 
and dynamical time 
($H_\mathrm{0}$t$_\mathrm{d}$ $\propto$ H$_\mathrm{0}$r$_\mathrm{ave}$/$\sigma$$_\mathrm{v}$).   
To compute these quantities we use r$_\mathrm{ave}$ instead 
of the harmonic radius r$_\mathrm{h}$, because 
we select groups according to their maximal extension rather than constraining 
their maximum galaxy-galaxy separation.
In  Figs.\,5 and  6 distributions of estimated M/L and 
H$_\mathrm{0}$t$_\mathrm{d}$ are shown. 
It appears that Ms possibly display higher M/L and shorter 
H$_\mathrm{0}$t$_\mathrm{d}$ than Ts, even though  
differences concerning these quantities are only marginally significant. 
The use of the harmonic radius (or of the median galaxy-galaxy separation) 
to compute these quantities  
would confirm the possible difference, 
with significance similar to that obtained 
with r$_\mathrm{ave}$.  
The higher mean M/L associated with Ms could indicate either a higher mean 
(M/L)$_\mathrm{gal}$ or a higher fraction of mass between 
galaxies.  
Concerning H$_\mathrm{0}$t$_\mathrm{d}$, the longer values associated 
with Ts might indicate that these are systems closer to turnaround, 
which are therefore less likely to be virialized.
Alternatively the smaller M/L and higher H$_\mathrm{0}$t$_\mathrm{d}$ 
associated with Ts might well be claimed to arise because of 
contamination by interlopers, and hence to be non-physical.  

In summary, the observed kinematical differences between Ts and Ms suggest 
that globally Ts do not constitute a fair subsample of Ms.   
Interestingly, differences are not significant between Ts and Ms in sample I, 
including mainly faint galaxies. 
\section {Simulated CG samples}
\begin{figure}
\resizebox{\hsize}{!}{\includegraphics{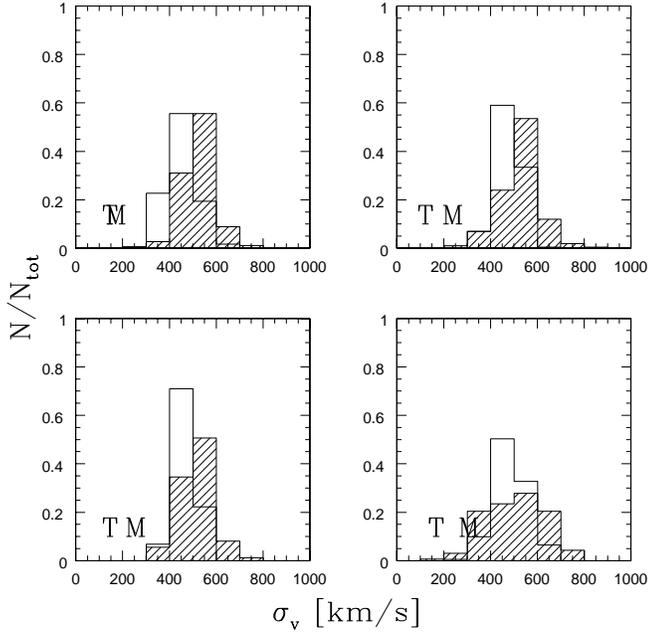}}
\hfill
\caption[]{ Distribution of the median $\sigma$$_\mathrm{v}$ in 
300 pseudo CG samples (Ts, solid histogram and Ms hatched) identified in 
simulated UZC catalogues in which radial velocity have been 
randomly reassigned. 
The median value of the real T and M samples are indicated by the 
symbols T and M. The real sample, which is 3-4 times as 
large as the simulated ones, appear to include many more low 
$\sigma$$_\mathrm{v}$ structures than reproducible by chance projection.} 
\end{figure} 
In order to probe physical reliability of kinematical differences 
between Ts and Ms, pseudo-CG samples must be produced by running 
the search algorithm on a large set of randomized UZC catalogues. 
This allows us to correctly evaluate how much of the 
kinematical differences between Ts and Ms might be attributed 
to random properties of the UZC galaxy distribution. 
Yet, randomly generated catalogues 
(i.e. random assignment of RA, Dec and cz within the catalogue limits) 
would completely destroy  
large-scale structures in the nearby universe and hence would not constitute 
fair comparison samples. 
Random reassignment of UZC galaxy coordinates (including redshift) 
leads to more realistic representations. 

In particular, to account for selection effects contaminating the  
velocity dispersion of T and M samples we have 
run the algorithm on 300 simulated UZC samples in which only the 
radial velocity of the galaxies has been reassigned. 
This leads to 
samples of $\approx$ 90 CGs (median=95, 87 and 101 first and last 
quartile) and allows to reproduce separately structures on the 
(projected) sky and in redshift space.  
Median values of the velocity dispersion distribution in the 300  
pseudo-CGs samples are displayed in fig.\,7, together with the median of 
samples of the real Ts (T) and Ms (M). 
It is evident that 
pseudo-CGs generally display   
$\sigma$$_\mathrm{v}$ larger than observed in the real sample and 
that they are unable to reproduce 
the severe segregation observed between Ts and Ms. 
Accordingly, random properties 
do not account for the much lower $\sigma$$_\mathrm{v}$ associated with Ts.  
Specifically, simulations indicate that for CGs between 2500 and 7500 
km/s, 
$\sigma$$_\mathrm{v}$ increases as n$^\mathrm{0.2}$. 
Subtracting this contribution from the observed slope 
yields the true increase in 
$\sigma$$_\mathrm{v}$ with multiplicity, 
which turns out to be $\sigma$$_\mathrm{v}$ $\propto$ n$^\mathrm{1.2}$. 
Accounting for field interlopers, which should bias the velocity 
dispersion of Ts towards the low end, only slightly reduces the steep slope 
in $\sigma$$_\mathrm{v}$. 
Indeed, rejection of systems with 
$\sigma$$_\mathrm{v}$$<$100 km\,s$^\mathrm{-1}$ 
yields (after correcting for random contributions) 
$\sigma$$_\mathrm{v}$ $\propto$ n$^\mathrm{0.9}$. 

Simulations which keep the projected position of galaxies 
are unable to fairly account for random properties affecting the 
average dimension of CGs. Therefore additional simulations 
have been run, in which RA and Dec of UZC galaxies have been 
separately reassigned. 
In this kind of simulated catalogues an average of 15 CGs are retrieved.  
The increase of CGs average dimension 
with multiplicity turns out to be rather modest ($\propto$ n$^\mathrm{0.2}$). 
Subtraction of this contribution from the observed one gives the 
correct increase of r$_\mathrm{ave}$ with n ($\propto$ n$^\mathrm{0.4}$).  
The space-number density of CGs thus appears to slightly decrease   
($\rho$ $\propto$ n$^{-0.2}$) from Ts to Ms, a trend which is not consistent 
with the relation expected in constant space-number density structures. 
Conversely, a small increase in surface number density 
($\Sigma$ $\propto$ n$^{0.2}$) holds, which might be induced by our 
request for a common projected scale for CGs of different multiplicity.
\section {Emission properties of galaxies in CGs }
UZC labels homogeneously the spectral classification (E=emission  
lines, A=absorption lines, B=E+A) for each galaxy, thereby allowing a  
check for possible links between emission properties and membership in 
CGs. 

To test whether samples of Ts and Ms are intrinsically different,  
the fraction of emission (with or without absorption lines) to absorption galaxies can be compared. 
This fraction also represents a rough estimate of the incidence of young 
(or rejuvenated) over old  objects, or alternatively of Spirals over Ellipticals. 
 Figure\,8 shows the Emission over Absorption (E/A) galaxy ratio for Ts 
(triangles) and Ms (squares) within each distance class. 
It is worth underlying that points in Fig. 8 indicate the ratio of the 
total population of emission galaxies over A galaxies in Ts and Ms. 

It emerges that the fraction of emission over absorption galaxies 
decreases from sample I to IV. 
This trend towards a larger fraction 
of galaxies with emission spectra increasing for lower galaxy luminosities 
was already known to exist both in the optical 
\citep{Zucca,Ratcliffe,Tresse} and in the near-IR \citep{Mamon01}. 
Any comparison of the emission line galaxy fraction 
with respect to kinematical parameters has to account for this trend 
which, concerning morphology, was already reported by Tikhonov (1990), 
Mamon (1990) and by Whitmore (1992). 
However, Fig. 8 shows that, even when accounting for the decrease 
of emission line galaxies with redshift, Ts include higher fractions 
of emission line galaxies than Ms. 
The luminosity of Ts and Ms member galaxies being similar, 
the trend of increasing fraction of emission-line galaxies with 
decreasing multiplicity is probably real. 
Galaxies in Ts and Ms in sample I  
display no significant differences, in accordance with kinematical 
similarities between Ts and Ms in this subsamples.   

Given that emission line galaxies are typically field galaxies, the data 
clearly suggest that Ts are more likely than Ms to be field 
structures (or to be contaminated by field interlopers) 
as already indicated by their lower $\sigma$$_\mathrm{v}$.    
To make this point more evident Fig.\,8 additionally displays 
the E/A ratio for Single galaxies 
and for galaxies in CGs which are ACO subclumps (ACO$_\mathrm{CG}$). 
Single galaxies are UZC galaxies which turn out to have no UZC 
companion(s) within an area of 200$h^\mathrm{-1}$kpc radius, 
and within $\Delta$cz$=\pm$1000\, km\,s$^\mathrm{-1}$ and form a plausible 
comparison sample for CGs on small scales. Among UZC galaxies  
single galaxies are $\approx$ 10 times more numerous than CG galaxies. 
It clearly emerges that CGs, whatever their luminosity, are lacking in 
gas rich galaxies when compared to single galaxies, 
and that the deficiency is larger for Ms. 
At the same time Fig.\,8 shows that 
CGs as a whole display an excess of spiral-rich galaxies 
when compared to those CGs which have been 
excluded from the sample because they turned out to be ACO$_\mathrm{CG}$. 
\begin{figure}
\resizebox{\hsize}{!}{\includegraphics{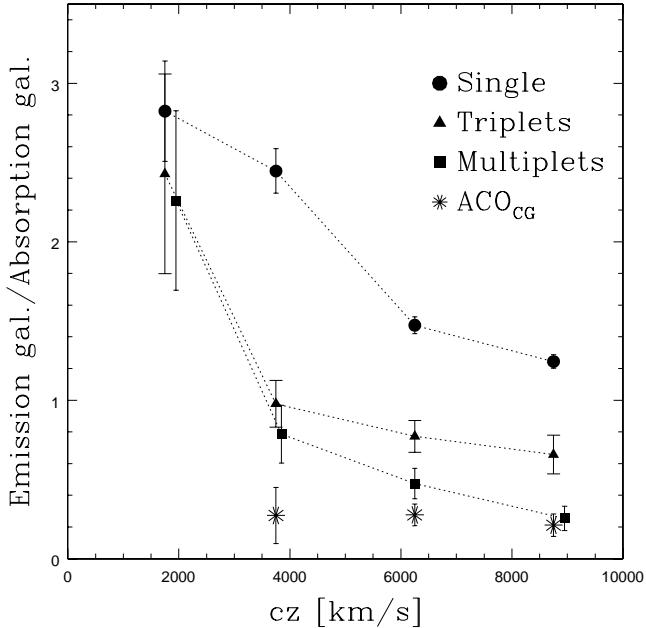}}
\hfill
\caption[]{Relative fraction of galaxies displaying Emission spectra 
over galaxies displaying Absorption spectra for Ts, Ms, for single 
galaxies in UZC (no neighbours within r=200$h^\mathrm{-1}$kpc)   
and for CGs that are ACO subclumps, which have been 
excluded from the main CG sample.}
\end{figure}  
\begin{figure}
\resizebox{\hsize}{!}{\includegraphics{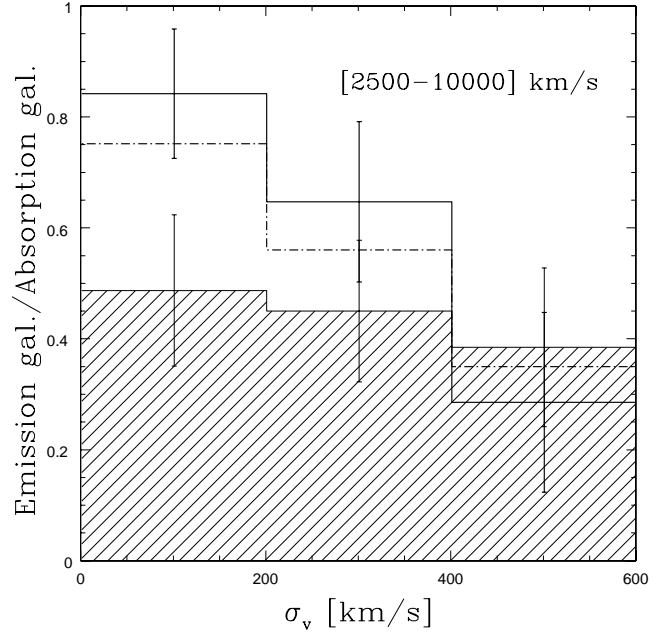}}
\hfill
\caption[]{Distribution of emission over absorption galaxy content 
as a function of CGs velocity dispersion. While in Ms (hatched area) 
the galaxy content is only modestly related to the velocity dispersion, 
the fraction of emission galaxies in Ts (solid histogram) turns out 
to be a strongly decreasing function of velocity dispersion. 
The hatched line shows the distribution for the whole CG 
sample. }
\end{figure}

Our data show the existence of a trend from single galaxies to 
galaxies in cluster subclumps, in which CGs occupy an intermediate 
position. Figure\,9, displaying the ratio of emission over absorption 
galaxies (in CGs at distance between 2500 and 10000 km/s) 
as a function of CG  $\sigma$$_\mathrm{v}$ confirms that a 
morphology-velocity dispersion  relation 
holds for the whole sample (hatched line), but also that the trend is  
induced by the inclusion among the CG sample of Ts (bold line) 
and specifically of low $\sigma$$_\mathrm{v}$ Ts. 
Accordingly, any process linking the increase of $\sigma$$_\mathrm{v}$ to 
the evolution of the spectral content of CGs is expected to be relevant 
predominantly in low multiplicity CGs.  
It is worth pointing out that if most low $\sigma$$_\mathrm{v}$ Ts are 
non-real structures, the morphology-velocity dispersion relation is 
not retrieved. 
 
The morphology-velocity dispersion relation is similar to the morphology-density relation observed in 
clusters and loose groups (Dressler 1980, Postman \& Geller 1984, Whitmore \& Gilmore 1991) 
with the fraction of gas-rich galaxies 
being a strong signature of multiplicity. The morphology-density 
relation has previously been shown to hold for HCGs 
(Mamon 1986, Hickson et al. 1988) with an offset relative to the general 
Postman \& Geller relation, indicating that at given spiral fraction, 
compact groups appear denser. 
It might be the inclusion 
within the sample of several spiral-rich, low multiplicity CGs 
that induces the offset, given that we find Ts to be even denser than Ms. 
Again, as for the morphology-velocity dispersion relation, 
the offset is to be reduced if most spiral rich, low $\sigma$$_\mathrm{v}$ 
Ts are non-physical systems.   

If the lower fraction of emission line galaxies in Ms corresponds to 
a lower fraction of Spirals, one accordingly expects the median 
(M/L)$_\mathrm{gal}$ of Ms members to be higher than for Ts galaxies. 
This could at least partially account for the higher M/L associated with Ms, 
although it remains uncertain whether the higher 
$\sigma$$_\mathrm{v}$ and  
early type galaxy content associated with Ms do indeed indicate that these are 
systems more evolved than Ts.   
Multiplicity also appears to strongly influences the behaviour of systems 
in Hickson's sample. Specifically we have shown \citep{Focardi} 
that the observed correlation between morphology and velocity 
dispersion in HCGs, \citep{Hickson88,Hickson92,Prandoni}    
just strongly reflects the different dynamical properties of systems 
with different multiplicity. 
 
In summary spectral characteristics indicate that two factors 
tend to strongly influence the number of emission line galaxies 
that will be retrieved in a CG sample. 
One is the fraction of faint galaxies included in the sample, with fainter 
galaxies being more likely to display emission line spectra. 
The second is the minimum multiplicity of CGs.    
The inclusion of Ts strongly biases a sample towards emission 
spectra galaxies. Combined with the average lower $\sigma$$_\mathrm{v}$, 
interactions between galaxies in Ts are accordingly predicted to be 
more disruptive than those in Ms, 
which suggests that perturbation patterns and/or asymmetric rotation 
curves \citep{Rubin} should be more frequent among Ts. 
\section {Large scale environment of Compact Groups}
Many embedded CGs are expected to be chance alignments of galaxies, 
not directly bound to one another along the line of sight,  
that form and destroy continuously within loose groups, whilst  
isolated CGs are generally assumed to be close dynamical systems, 
whose future evolution is a function of internal parameters only. 
Unfortunately, defining a CG as isolated is a non trivial problem, 
as one has to define boundaries (in space and luminosity) below which 
external galaxies perturb CG evolution and above which perturbations 
are negligible. 
Previous studies of CG environments yield contradictory results. Rubin et 
al. (1991) studying 21 HCGs find Ts to be more isolated than Ms.  On the 
other hand Barton et al. (1996) do not confirm this result. 
However hardly any isolated CG should be retrieved, as bright galaxies  
are known to be strongly clustered, and faint galaxies are known to 
cluster around bright ones \citep{Benoist,Cappi}.
 
In order to properly investigate possible 
relations between small and large scale environments, the algorithm counts 
neighbours (N$_\mathrm{env}$) for each CG within 
a distance $\Delta$R =1 $h^\mathrm{-1}$Mpc and 
$\Delta$v$^\mathrm{II}$=$\pm$1000\,km\,s$^\mathrm{-1}$ from the CG center. 
\begin{figure}
\resizebox{\hsize}{!}{\includegraphics{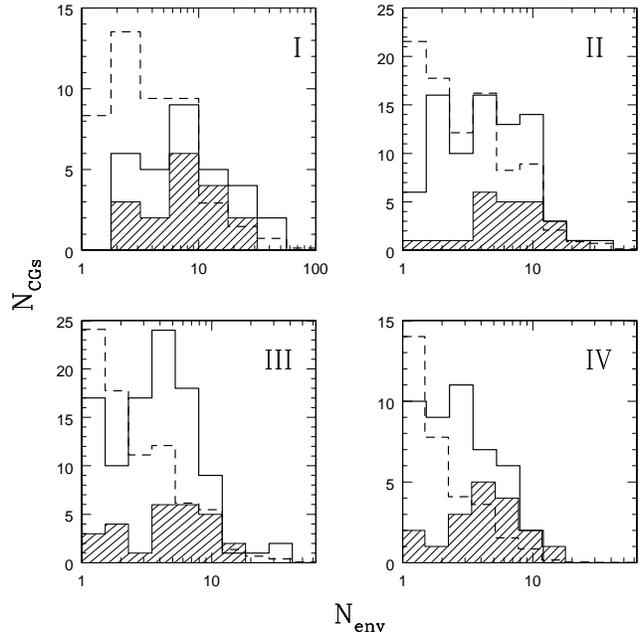}}
\hfill
\caption[]{CGs distribution as a function of large scale neighbours. 
Neighbours have been counted out to 
R=1$h^\mathrm{-1}$\,Mpc and $\Delta$cz$=$$\pm$1000 from 
the CG center. The solid histogram refers to Ts, the hatched one to Ms. 
The dashed line shows the large scale environment distribution 
of single galaxies (on a scale that fits into the limits set by CGs).}
\end{figure} 
To minimize distance uncertainties due to the relative 
incidence of peculiar motions neighbourhood richness is computed only 
for CGs at cz$>$1500 km\,s$^\mathrm{-1}$, thereby reducing the number of  
Ts and Ms in subsample I from 34 to 31 and from 21 to 17 respectively.  
In  Fig.\,10 the overall distribution of CGs with respect 
to N$_\mathrm{env}$ is shown. 
The solid line refers to Ts, the hatched area to Ms.  
It clearly emerges that Ts are more likely than Ms to be found 
in isolated environments, but that, compared to the much more numerous 
single galaxies (hatched line), their environment is denser. 
According to the KS test differences between Ts and Ms 
are significant at 98\%, 97\% and 94\% c.l. in subsamples II, 
III and IV, whilst they are non significant in class I. 
In simulated samples Ms show no excess of neighbours 
with respect to Ts, so that no corrections for selection effects 
have to be applied to the environmental data.   
Thus we find three independent parameters 
(velocity dispersion, spectral properties and environmental 
density) suggesting that Ts and Ms constitute different populations. 
The sparser environment, the higher emission line fraction and the lower 
velocity dispersion of Ts all might result from  
high contamination by field interlopers. 
However they are also compatible with Ts being recently formed systems 
of field galaxies, not yet embedded within a common virialized halo,   
in which dynamical friction efficiently 
transfers orbital energy of the group into the internal energy of a 
single merger remnant. 
   
It must also be stressed that although the probability of chance 
alignments decreases rapidly when going from Ts to Ms, the richer 
environment associated with Ms enhances the probability of Ms being chance 
alignments. Indeed, simulations indicate (Mamon, private communication) 
that a Multiplet with 5 neighbours is twice 
as likely to be a chance alignment than a 
Triplet with less than 2 neighbours.  
Finally, we underline that the relation between  
CGs multiplicity  
and large scale galaxy density indicates that when isolation  
is used as a CG selection parameter the sample 
is biased towards either luminous or low multiplicity CGs, 
the former including many early type galaxies, the latter a high fraction of 
late type galaxies. The requirement of isolation consequently induces 
large scatter in the spectral (and morphological) properties of CG samples. 
\section{Surface density contrast}
Next we combine information about multiplicity and  
environment and compute the surface density contrast 
of CGs ($\Sigma$$_\mathrm{CG}$/$\Sigma$$_\mathrm{env}$), 
a parameter that quantifies the excess 
of surface density within the CGs as compared to that of their environment. 
\begin{figure}
\resizebox{\hsize}{!}{\includegraphics{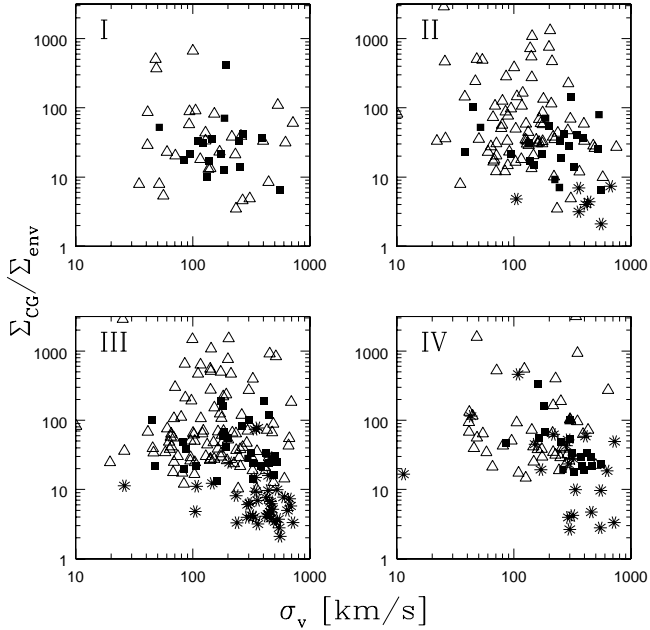}}
\hfill
\caption[]{Surface density contrast of Ts (empty triangles) and Ms (filled squares) versus velocity dispersion  $\sigma$$_\mathrm{v}$. 
Stars represent ACO$_\mathrm{CGs}$. The diagram clearly shows that CGs with 
different environmental and dynamical properties can be separated according to multiplicity. }
\end{figure} 
It is defined as the surface number density of galaxies within 
and area of radius r$_\mathrm{max}$ with respect to the surface number 
density of galaxies within an area of radius 1$h^\mathrm{-1}$Mpc. 
A space number density contrast 
constraint (defined with respect to the mean of the entire sample), 
has previously been coupled to the FoF algorithm to identify 
loose galaxy groups in the CFA \citep{Ramella} and SRSS 
\citep{Maia} surveys. 
If we were to adopt a similar criterion, 
CGs would correspond to even higher overdensities, 
because the vast majority of the systems defined on a 
200$h^\mathrm{-1}$ kpc scale 
turn out to be single galaxies, which, as shown in figure 10, are associated 
with environments typically sparser than those of CGs. 
  
If surface density contrast is plotted against $\sigma$$_\mathrm{v}$  
one expects field systems to occupy low velocity dispersion-high density 
contrast regions and systems which are subclumps embedded within larger 
structures to occupy high velocity dispersion-low density contrast regions. 
In  Fig.\,11 the region occupied by CGs in a 
$\Sigma$$_\mathrm{CG}$/$\Sigma$$_\mathrm{env}$ vs. $\sigma$$_\mathrm{v}$ 
plot is displayed. 
Whilst Ts are located predominantly near the field-systems area, Ms are 
typically associated with the embedded-systems area. 
Figure 11 shows that multiplicity is a rather robust parameter 
to discriminate between field structures and embedded structures, 
and indicates that, to reduce scatter in CG properties,  
Ts should not be included among higher multiplicity CGs, as this roughly 
would correspond to sampling together field-CGs and embedded-CGs. 

In Fig.\,11 the CGs that have been excluded from the
sample because they are ACO subclumps are also plotted. ACO 
subclumps occupy a distinct region on the diagram. 
Whilst presenting a velocity dispersion similar to Ms, 
ACO$_\mathrm{CGs}$ are generally less overdense structures. 
On one side this might confirm that several Ms are structures that 
constitute the central core of large-groups/poor-clusters. 
This interpretation nicely matches observations indicating that, 
concerning X-ray properties, the distinction between compact and loose 
groups is not a fundamental one \citep{Heldson}. 
On the other hand the embedded status of many Ms could indicate that these
are actually temporary chance alignments within a structure much larger than
the CG \citep{Mamon86,Walke,Hernquist}. 
If this is the case, the characteristic $\sigma$$_\mathrm{v}$ 
associated with Ms are probably too high an estimate, and all dynamically 
derived parameters, such as $\mathrm{M/L}$ or the dynamical 
time H$_\mathrm{0}$t$_\mathrm{d}$ would strongly reflect the same bias. 

To underline that the properties and differences between Ts and Ms are 
not to be attributed to random properties of the large scale distribution 
of UZC galaxies we show in  Fig.\,12 the position occupied by  
pseudo-CG samples extracted from simulated UZC catalogues (see section 5) on 
a $\Sigma$$_\mathrm{CG}$/$\Sigma$$_\mathrm{env}$ vs. 
$\sigma$$_\mathrm{v}$ plot. 
As pseudo-CG samples typically include few systems, 
to match the numerical dimension of the real sample we have grouped 
together 20 pseudo-CG samples.
\begin{figure}
\resizebox{\hsize}{!}{\includegraphics{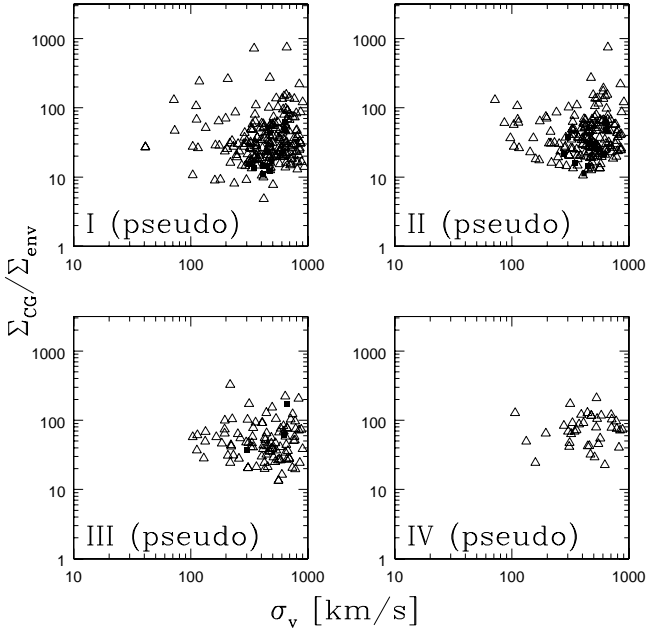}}
\hfill
\caption[]{Surface density contrast versus $\sigma$$_\mathrm{v}$ in pseudo 
CG samples. To achieve numerical consistency with the real CG sample 20   
pseudo-samples have been grouped together. No relation appears to exist  
between $\sigma$$_\mathrm{v}$ and surface density contrast in CG samples 
identified in random UZC catalogs. }
\end{figure}

\section{Sample selection criteria and CGs properties}
The different kinematical, morphological and environmental 
behaviour displayed by Ts and Ms allows us to relate   
commonly adopted 
CG selection criteria to the sample's properties. 

Specifically our analysis has shown that multiplicity, velocity dispersion, 
large scale galaxy density and spectral/morphological mix are strongly 
linked together. 
Even if the explicit CG selection criterion constrains only one of these 
parameters, the link between them causes the remaining selection parameters 
to be constrained too. 
A constraint on compactness, such as the one we have adopted,    
 biases a sample towards low multiplicity 
structures and thus indirectly towards low velocity dispersion, spiral 
rich, isolated structures. A small limit on the maximum velocity 
dispersion of CG members acts in the same direction.  
Therefore a parallel requirement for isolation though non affecting the 
low multiplicity CGs will severely reduce the number of high 
multiplicity CGs. Conversely, requiring a minimum of four members will 
bias a CG sample towards intrinsically embedded, spiral-poor groups. 

Our analysis shows that Ms, whose parameters are statistically more reliable,
happen to be more likely to constitute embedded subclumps 
whilst Ts might be more likely to be contaminated by 
systems which are just unbound projections of field galaxies.  
Interestingly, disregarding CGs in which faint members are 
counted (sample I), extremely compact systems with a minimum of 4 members 
which also fulfill the isolation criterion appear to be extremely rare indeed, 
thereby fitting predictions by numerical simulations claiming that 
compact configurations are rapidly destroyed \citep{Mamon87,Barnes}.    

Provided the fraction of non-physical CGs does not dominate 
the statistics, a large scatter in CG properties results when the analysis 
does not distinguish between Ts and Ms, as multiplicity 
appears to be a preliminary, robust discriminant between less evolved,  
field-systems and more evolved, embedded systems.     
Concerning HCGs, the suggestion that high and low $\sigma$$_\mathrm{v}$ 
groups are intrinsically distinct can already be found in Mamon (2000), 
who further states that low $\sigma$$_\mathrm{v}$ groups are either 
chance alignments or systems in their final stages of coalescence. 
The new point we add here is that low $\sigma$$_\mathrm{v}$ systems are 
typically low multiplicity field structures.  

That HCGs constitute a heterogeneous sample has previously also been stressed 
by de Carvalho et al. (1997) and Ribeiro et al. (1998), who, 
based on the analysis of 17 HCGs, identify 3 distinct CG families.  
They suggest that these correspond to  
3 different dynamical stages, specifically they interpret embedded CGs 
as precursors of isolated and very dense systems.   
In comparison with Ribeiro et al. (1998) and based on our much larger 
CG sample we interpret low velocity dispersion, high overdense CGs 
(mostly Ts occurring along low density filaments) as the bottom level of 
the clustering process and embedded structures (either chance 
projections or collapsing cores within loose groups/poor clusters) 
such as systems in a more advanced evolutionary stage. 
Our interpretation, which explains the 
weak X-ray emission of field CGs in terms of their shallow potential 
wells \citep{Heldson}, requires that when X-ray emission is observed 
in small, gas rich CGs, it 
should be totally ascribable either to individual galaxies or 
to collisional shock-heating of the gas in low luminosity systems. 

Our analysis indicates that interactions should be efficient mainly  
in the most overdense, low velocity dispersion structures, 
which are mainly Ts that include high fractions of gas-rich galaxies.   
Accordingly, it is not surprising that statistical analysis looking for 
interaction in HCGs (which include many n$>$3 CGs embedded within a 
common halo) 
globally reveals low fractions of merging remnants and 
blue Ellipticals \citep{Zepf}.   
Actually, the suggestion that disturbances should be enhanced 
only among Ts better fits observations 
reporting that the most easily detected disturbed galaxies are spirals 
in small groups \citep{Fried} and that the most spectacular mergers, 
such as bright IRAS galaxies (ULIRGs), 
appear to involve strong interactions of gas-rich galaxies where the 
pairs are either isolated or part of small groups \citep{Sanders}.  
It is also worth pointing out that the request for a minimum of 4 members 
which has biased the HCGs towards intrinsically embedded, gas-poor member 
groups, possibly explains why, despite the high expected interaction rate, 
HCGs as a whole present rather low evidence for strong AGN-starbursting 
episodes \citep{Coziol1998,Kelm,Coziol2000}.    

Whether kinematical differences between Ts and Ms 
are generally compatible with hierarchical model predictions depends 
upon the specific assumptions one makes on the (M/L)$_\mathrm{gal}$ 
of Ts and Ms member galaxies and on the fractional group 
mass (f$_\mathrm{gal}$) associated with its galaxies. 
Provided the luminosity of CG members is independent of 
multiplicity, and assuming (M/L)$_\mathrm{gal}$ and f$_\mathrm{gal}$ 
is the same for Ts and Ms, one predicts 
r$_\mathrm{ave}$\,$\propto$\,N$^\mathrm{1/3}$ 
and  $\sigma$$_\mathrm{v}$\,$\propto$\,N$^\mathrm{1/3}$.  
While the r$_\mathrm{ave}$ slope is roughly consistent with 
these expectations, 
the $\sigma$$_\mathrm{v}$ slope increases much faster. 
Indeed, figure 8 suggests that the assumption 
concerning the same (M/L)$_\mathrm{gal}$ for Ts and Ms is probably 
not satisfied, as absorption and emission galaxies are expected to represent 
ellipticals and spirals, 
and the former are typically associated with higher (M/L) galaxies than the 
latter.   
While (M/L)$_\mathrm{gal}$ is expected to increase with 
multiplicity,  f$_\mathrm{gal}$ might actually decrease, 
due to the fact that higher multiplicity CGs are more likely to be 
associated with gas-rich, X-emitting groups.  
Consequently, before assessing whether globally data on CGs  
are (or are not) compatible with 
hierarchical model predictions, 
more accurate models, taking into account 
the different (M/L)$_\mathrm{gal}$ 
and f$_\mathrm{gal}$ of Ts and Ms, should be investigated. 

\section{Conclusions}
We have taken advantage of a large, almost complete 3D catalogue 
to identify a sample of 291 northern CGs with redshift between 1000 
and 10\,000 km\,s$^\mathrm{-1}$. CGs include a minimum of 3 members which 
have to lie within a region of 200$h^\mathrm{-1}$ kpc and 
$\Delta$v=$\pm$1000 km\,s$^\mathrm{-1}$. 
Kinematical properties of CGs and spectral characteristics of member 
galaxies have been investigated and related to large-scale environmental 
parameters. 
The sample has been used to compare Triplets, 
which constitute 76\% of the sample, to higher-multiplicity structures. 
The analysis indicates that multiplicity is intrinsically linked to CG 
properties such as velocity dispersion, large-scale environment and 
spectral characteristics of galaxies.  
 
Specifically, it is found that Ts are more likely to be isolated systems and  
to display low velocity dispersion as well as a high gas-rich galaxy content.  
We suggest that Ts, although affected by interlopers,   
generally correspond to field galaxy structures. 
They constitute ideal sites 
for efficient merging to occur, and are thus likely to 
transform into a single galaxy 
as continuous accretion from surrounding galaxies is not viable on times 
shorter than their dynamical time scale.   

On the other hand, higher multiplicity CGs are mainly associated with  
high velocity dispersion systems, 
whose members are preferentially gas-poor galaxies.    
These CGs display lower density contrast than field CGs and may thus suffer 
contamination by systems that are just temporary chance alignments within 
loose groups/poor clusters. Those Ms which are real physical systems 
should constitute the center of a larger collapsing group and are thus 
expected to display diffuse X-ray emission.  

In summary, our data indicate that, provided most CGs are real physical 
systems, Ts and Ms correspond to two extremely 
different classes of systems. Therefore, any fair analysis of CGs properties 
should treat Ms and Ts separately. 
\begin{acknowledgements}
We are pleased to thank S.\,Bardelli, A.\,Cappi, S.\,Giovanardi, P. Hickson, 
A.\,Iovino, G.G.C.\,Palumbo, E.\,Rossetti, R.\,Sancisi, G.\,Stirpe and V.\,Zitelli 
for stimulating discussions and suggestions. 
We thank the referee G.\,Mamon whose comments and criticism greatly 
improved the scientific content of the paper. 
This work was supported by MURST. B.K acknowledges a fellowship of  
Bologna University. 
\end {acknowledgements}

\end{document}